\documentclass{article}

\usepackage{PRIMEarxiv}

\usepackage[utf8]{inputenc} 
\usepackage[T1]{fontenc}    
\usepackage{hyperref}       
\usepackage{url}            
\usepackage{booktabs}       
\usepackage{amsmath}
\usepackage{amsthm}
\usepackage{amssymb}
\usepackage{amsfonts}
\usepackage{mathtools}
\usepackage{amsfonts}       
\usepackage{nicefrac}       
\usepackage{microtype}      
\usepackage{fancyhdr}       
\usepackage{graphicx}       
\graphicspath{{media/}}     
\usepackage{times}
\usepackage{subfig}
\usepackage{xcolor}
\usepackage{array}
\usepackage{cite}
\usepackage{multirow}
\usepackage{adjustbox}
\usepackage[scr=dutchcal]{mathalfa}
\usepackage[font=small,labelfont=bf]{caption}
\usepackage{algorithm}
\usepackage{algpseudocode}
\usepackage{enumitem}

\DeclareMathOperator{\diag}{diag}

\newcommand{\greencheck}{{\color{green}\checkmark}}
\newcommand{\redtimes}{{\color{red}\times}}

\newtheorem{theorem}{Theorem}
\newtheorem{lemma}{Lemma}

\newtheorem{definition}{Definition}

\newtheorem{remark}{Remark}

\usepackage{tikz-network}
\tikzset{
    block/.style = {draw, fill=white, rectangle, minimum height=.75cm, minimum width=1.25cm},
    tmp/.style  = {coordinate}, 
    sum/.style= {draw, fill=white, circle},
    input/.style = {coordinate},
    output/.style= {coordinate},
    pinstyle/.style = {pin edge={to-,thin,black}
    }
}

\title{On the Comparison of Reinforcement Learning and Adaptive Control for Linear Systems under \\
Packet Loss and Uncertainty
}

\author{
  Moh. Kamalul Wafi\\
  Department of Electrical \& Computer Engineering \\
  Northeastern University \\
  Boston, MA 02115 USA \\
  \texttt{\{wafi.m\}@northeastern.edu} \\
}

\begin{document}
\maketitle

\begin{abstract}
This paper presents a comparative study between Adaptive Quantized Control (AQC) and Deep Deterministic Policy Gradient (DDPG) reinforcement learning for uncertain linear systems with input quantization over communication channels subject to packet loss. The considered setting also includes dynamic switching from a nominal unstable system to a more unstable one during operation. The AQC is designed for unknown system dynamics using acknowledgment messages to compensate for packet losses, whereas the DDPG controller is trained using the nominal system model without acknowledgment messages. Numerical results show that the DDPG controller achieves faster transient responses and improved damping within its training environment. However, under model uncertainty, packet loss, and dynamic switching, the AQC consistently demonstrates superior robustness owing to its rigorous Lyapunov stability guarantees. These results highlight the trade-off between data-driven performance and model-based robustness, and provide insight into the applicability of reinforcement learning and adaptive control for networked uncertain systems.
\end{abstract}

\allowdisplaybreaks

\section{Introduction}\label{S1}
Modern systems' complexity has presented several challenges for classical and modern control techniques. However, there has been a recent trend toward utilizing learning-based control methods, particularly in high-dimensional spaces \cite{R1}. Over the past decade, deep learning has significantly impacted both the theoretical and practical aspects of machine learning by enabling its application to non-Euclidean spaces, such as graphs with interdependencies. This revolution has expanded the scope and capabilities of machine learning beyond traditional Euclidean spaces \cite{R2}. The integration of deep learning and reinforcement learning, known as deep reinforcement learning (DRL), has been the subject of extensive control-related studies \cite{R3,R4,R7,Wafi-ACC24}. The ideas behind DRL are similar to those of deep Q-network (DQN) as seen in \cite{R10}. The DQN is limited to discrete and finite actions whereas DRL, on the other hand, allows for continuous and infinite control actions, as described in \cite{R11}. In this work, we analyze the behavior of an unstable system under two different types of control: the deep deterministic policy gradient (DDPG) reinforcement learning, which is learning-oriented and the quantized control which is adaptive-oriented.

Additionally, the term ``quantization'' in this context refers to the limitation of communication within a specific bandwidth in networked systems. The concept of using quantization for stabilization of linear systems with finite control signals and measurements was introduced in the reference \cite{R12}. The state information is quantized in a coarse manner, with the level of precision becoming finer as it approaches the origin in a logarithmic manner. This can also be alternatively explained using the more widely accepted sector-bounded quantizer  \cite{R13}. Beyond that, dealing with the uncertain system, the adaptive control with input quantizer is studied by \cite{R14} which is also implemented into systems with packet loss $\eta$ \cite{R15}. This means there exists a probability $\Bar{p}$ of control signal not being sent to the plant, with some variations of the non-linear uncertain system \cite{R16}. Similar communication and uncertainty issues also arise in networked estimation and process control applications, including hydraulic multi-tank systems, where reliable state estimation and feedback control are essential for maintaining system performance \cite{Wafi-QuadrupleTank,Wafi-ThreeTank}. Note that with stabilization guaranteed \cite{R12,R13,R14,R15}, it is interesting to see how the learning approach performs under some limitations \cite{R17}. 

The DDPG reinforcement learning is a combined deterministic-actor $a = \mu_\theta(s)$ instead of the stochastic $\pi_\theta(a|s) = \mathbb{P}[a|s;\theta]$ and the Q-value critic $Q(s,a)$ which both applies the feed-forward neural network (FFNN) \cite{R11,R18}. The stability is guaranteed as presented in \cite{R19} over uncertain systems with sector-bounded for the non-linear activation function. However, we design the dynamic changes in the simulation to see how robust the trained DRL control performs under specific trained system $A$ considering the dynamic change beyond the environment $A_w$. The two dynamics encompass systems ranging from modestly unstable to highly unstable, as defined later.

The remainder of this paper is organized as follows. Section~\ref{sec:aqc} reviews the adaptive quantized control framework with acknowledgment messages and packet loss. Section~\ref{sec:ddpg} presents the actor--critic DDPG reinforcement learning controller and its quantized implementation. Numerical comparisons between the two approaches are reported in Section~\ref{sec:results}, followed by concluding remarks in Section~\ref{sec:conclusion}. This article is an extended and refined version of our preliminary conference paper \cite{Wafi-SIAM}.

\begin{figure}[t!]
    \centering
    \includegraphics[width=.5\linewidth]{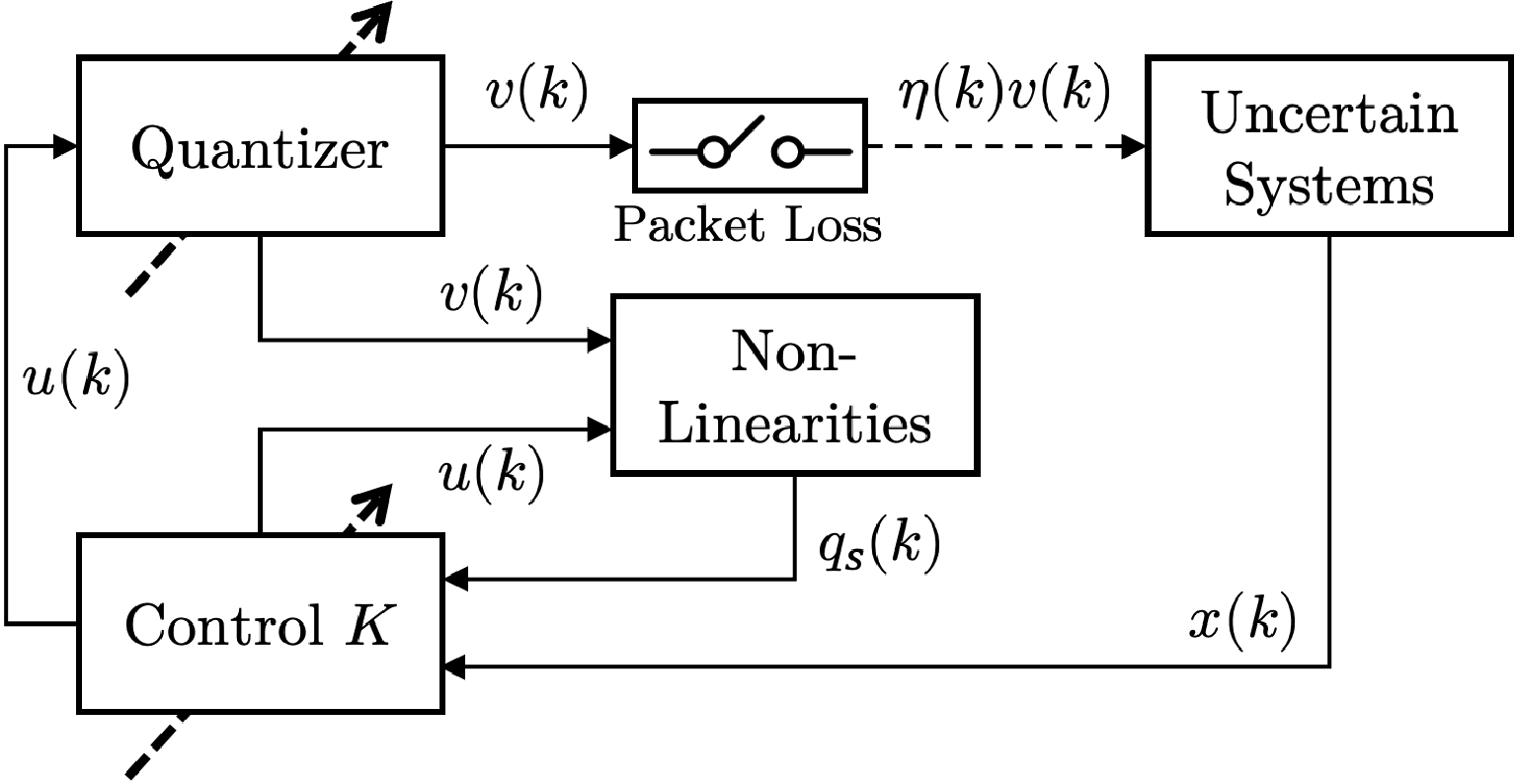}
    \caption{Adaptive quantized control ($K,v$) method with time-varying quantizer $q(k,u(k))$. The solid arrows between blocks mean the connected information whereas the dashed arrow show there exists a possibility not being connected. The dashed arrows behind blocks deduces the time-varying.}
    \label{F1}
\end{figure}

\section{Adaptive Quantized Control}\label{sec:aqc}

In this section, we briefly review the adaptive quantized control (AQC) framework proposed in \cite{R14} and extended to systems subject to packet loss in \cite{R15}. Consider the discrete-time uncertain linear system shown in Fig.~\ref{F1}, where the quantized control signal is transmitted through a communication channel subject to packet loss. 

The plant is described by
\begin{align}\label{eq:plant}
    x(k+1)=Ax(k)+\eta(k)Bv(k),
\end{align}
where $k\in\mathbb{N}_0$ is the discrete-time index and $x(0)=x_0$ is the initial condition. The vector $x(k)\in\mathbb{R}^n$ denotes the system state, $v(k)\in\mathbb{R}^m$ is the quantized control input, $A\in\mathbb{R}^{n\times n}$ is the unknown state matrix, and $B\in\mathbb{R}^{n\times m}$ is the known input matrix.

The binary random variable $\eta(k)\in\{0,1\}$ models packet transmission, where $\eta(k)=1$ indicates that the control packet is successfully received by the plant, whereas $\eta(k)=0$ indicates packet loss. We assume that $\{\eta(k)\}_{k\ge0}$ are independent and satisfy
\[
    \mathbb{P}\{\eta(k)=0\}\le\bar{p},
    \qquad \text{for every } k,
\]
where $\bar{p}\in[0,1)$ is an upper bound on the packet-loss probability.

The quantized control input is generated from the nominal control law
\begin{align}\label{eq:quantizer}
    v(k)=q(k,u(k)),
\end{align}
where
\[
    u(k)=H(k)x(k),
\]
and $H(k)\in\mathbb{R}^{m\times n}$ denotes the time-varying feedback gain. The quantizer $q(\cdot,\cdot)$ acts componentwise according to the logarithmic quantization rule
\begin{align}\label{eq:log_quantizer}
    q_i
    \triangleq
    \begin{cases}
    \varphi_i(k,j), &
    \textrm{if }u_i\in(\varphi_i(k,j+1),\varphi_i(k,j)],\\
    -\varphi_i(k,j), &
    \textrm{if }u_i\in[-\varphi_i(k,j),-\varphi_i(k,j+1)),\\
    0, &
    \textrm{if }u_i=0,
    \end{cases}
    \qquad
    j\in\mathbb{I},\;
    i=1,\ldots,m,
\end{align}
where
\[
    \varphi_i(k,j)=a_i(k)\rho_i^j(k), \qquad i=1,\ldots,m,
\]
with $a_i(k)>0$ and $0<\rho_i(k)<1$. Here, $a_i(k)$ determines the quantization scale, whereas $\rho_i(k)$ specifies the quantizer coarseness. The functions $q_i(\cdot,\cdot)$ and $u_i(\cdot)$ denote the $i$-th components of $q(\cdot,\cdot)$ and $u(\cdot)$, respectively. For negative indices, the quantization levels are defined by
\[
    \varphi_i(k,-j)\triangleq
    \frac{a_i(k)}{\rho_i^j(k)}, \qquad i=1,\ldots,m.
\]

\begin{figure}[t!]
    \centering
    \includegraphics[width=.65\linewidth]{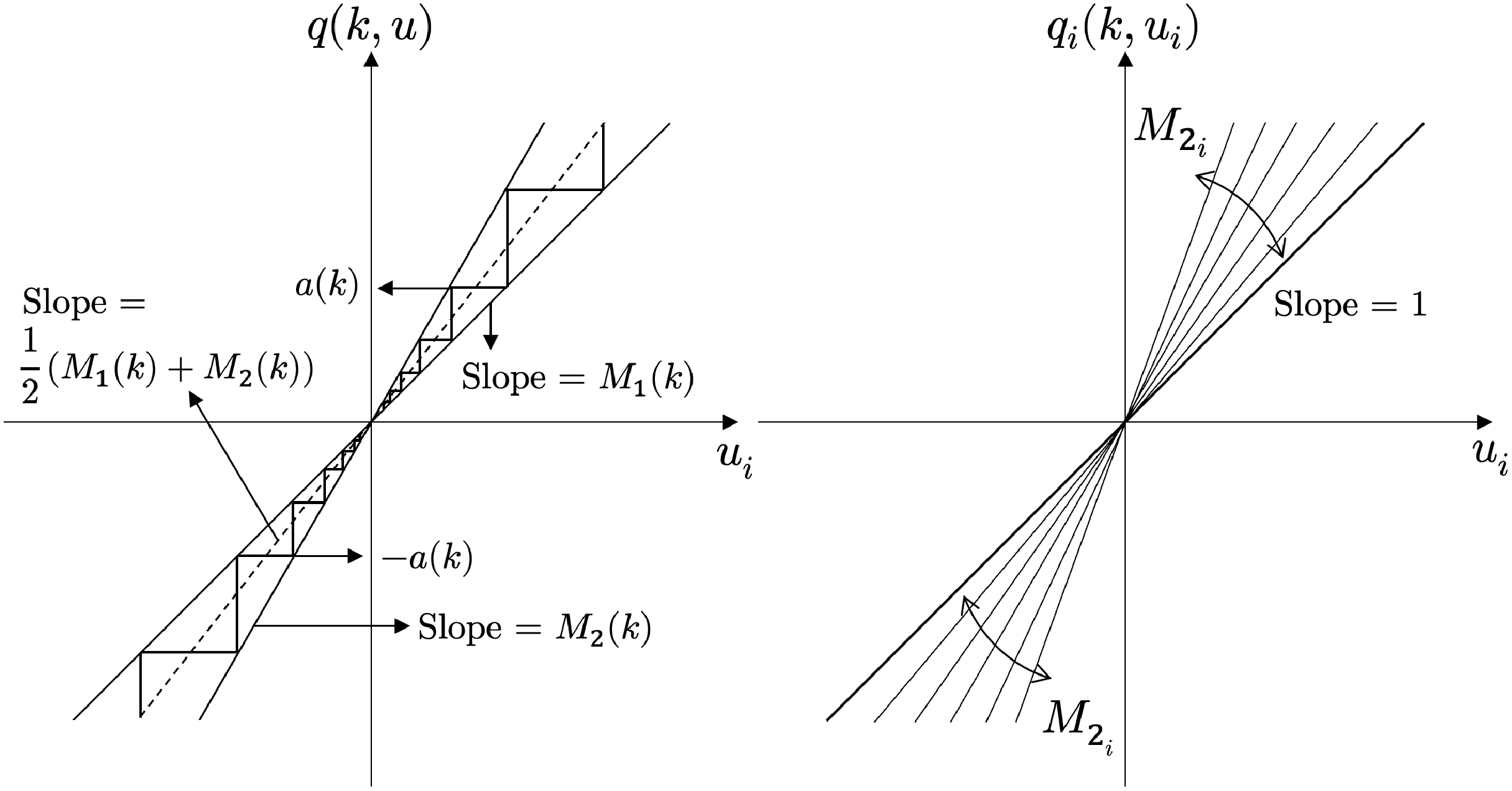}
    \caption{Left: Logarithmic quantizer $q$ for the scalar case ($m=1$). Right: Example of the corresponding sector bounds with $M_1(k)\equiv1$ and $M_2(k)\in\{1+\hat{a}\mu_i^j:j\in\mathbb{I}\}$.}
    \label{F2}
\end{figure}

The logarithmic quantizer in \eqref{eq:log_quantizer} admits an equivalent representation as a time-varying sector-bounded memoryless nonlinearity \cite{R13},
\begin{align}\label{eq:sector}
\begin{aligned}
    \mathcal{Q}
    \triangleq
    &\Big\{
    q:\mathbb{N}_0\times\mathbb{R}^m\rightarrow\mathbb{R}^m:
    \; q(\cdot,0)=0, \\
    &\left[q(k,u)-M_1(k)u\right]^\top
    \left[q(k,u)-M_2(k)u\right]
    \le0,\\
    &u\in\mathbb{R}^m,\; k\in\mathbb{N}_0\Big\}.
\end{aligned}
\end{align}
Here, $M_1\in\mathbb{R}^{m\times m}$ and $M_2\in\mathbb{R}^{m\times m}$ denote diagonal matrices given by
\begin{align*}
    M_1(k)&\triangleq\diag\!\left(M_{1_1}(k),\ldots,M_{1_m}(k)\right)\succ0, \\
    M_2(k)&\triangleq\diag\!\left(M_{2_1}(k),\ldots,M_{2_m}(k)\right)\succ0,
\end{align*}
which satisfy $M_2(k)-M_1(k)\succ0.$ The sector bounds satisfy
\[
    \rho_i(k)=\frac{M_{1_i}(k)}{M_{2_i}(k)},
    \qquad i=1,\ldots,m,
\]
and are illustrated in Fig.~\ref{F2}.
Equivalently, \eqref{eq:sector} can be written componentwise as
\begin{align}\label{eq:sector_scalar}
    M_{1_i}(k)u_i^2 \le q_i(k,u_i)u_i\le M_{2_i}(k)u_i^2,
\end{align}
for all $u_i\in\mathbb{R}$ and $k\in\mathbb{N}_0$. Furthermore,
\begin{align}\label{eq:rho}
    \rho_i(k)
    = \frac{M_{1_i}(k)}{M_{2_i}(k)}
    = \frac{1-\beta\delta_i(k)}
         {1+\beta\delta_i(k)},
    \qquad i=1,\ldots,m,
\end{align}
where
\[
    \delta(k)
    \triangleq
    \frac{1}{\beta}
    \left[M_2(k)+M_1(k)\right]^{-1}
    \left[M_2(k)-M_1(k)\right],
    \qquad \beta\neq0.
\]
Hence, $\delta(k)$ completely determines the quantizer coarseness.
Defining
\begin{align}\label{eq:delta}
    \Delta(k)
    &\triangleq
    \diag\!\left(
    \delta_1(k),\ldots,\delta_m(k)
    \right)
    \nonumber\\
    &=
    \frac{1}{\beta}
    \left[M_2(k)+M_1(k)\right]^{-1}
    \left[M_2(k)-M_1(k)\right],
\end{align}
the diagonal matrix $\Delta(k)$ will be used in the adaptive controller design presented next.

For completeness, we briefly summarize the adaptive quantized control framework and refer the reader to \cite{R14,R15} for the complete controller design and stability analysis. Following \cite{R14}, the quantizer is decomposed into linear and nonlinear components as
\begin{align}\label{eq:quantizer_decomp}
    q(k,u)
    =
    \frac{1}{2}\left[M_1(k)+M_2(k)\right]u
    +
    q_s(k,u),
\end{align}
where
$q_s:\mathbb{N}_0\times\mathbb{R}^m\rightarrow\mathbb{R}^m$
denotes the nonlinear component satisfying
\begin{align}\label{eq:sector_qs}
\begin{aligned}
    \mathcal{Q}_s
    \triangleq
    &\Big\{
    q_s:\mathbb{N}_0\times\mathbb{R}^m
    \rightarrow
    \mathbb{R}^m:
    \;q_s(\cdot,0)=0,\\
    &q_s^\top(k,u)q_s(k,u)
    -
    \frac{1}{4}
    u^\top
    \left[M_2(k)-M_1(k)\right]^2
    u\le0,\\
    &u\in\mathbb{R}^m,\;
    k\in\mathbb{N}_0
    \Big\}.
\end{aligned}
\end{align}
For the subsequent stability analysis, we adopt the notion of Lyapunov stability in probability from \cite{R20}.

\begin{definition}
Consider the stochastic discrete-time system
\begin{align}\label{eq:stochastic_system}
    x_{k+1}=f(x_k,y_{k+1}),
    \qquad
    k\in\mathbb{N}_0,
\end{align}
where $x_k\in\mathbb{R}^n$ and $\{y_k:k\in\mathbb{N}_0\}$ is an $\mathbb{R}^d$-valued stochastic process defined on the probability space $(\Omega,\mathbb{F},\mathbb{P})$. Here, $\Omega$ denotes the sample space, $\mathbb{F}$ the associated $\sigma$-algebra, and $\mathbb{P}:\mathbb{F}\rightarrow[0,1]$ the probability measure. The stochastic process $y_k$ is measurable and maps $\Omega$ into the state space $\Omega_0\subseteq\mathbb{R}^d$. Define
\[
    \mathbb{F}_k=\sigma(y_1,\ldots,y_k),
    \qquad
    k\ge1,
\]
with $\mathbb{F}_0=\{\emptyset,\Omega\},$
so that $\{\mathbb{F}_k\}_{k\ge0}$ forms an increasing filtration.
The origin of \eqref{eq:stochastic_system} is said to be
\begin{enumerate}[label=(\roman*)]
    \item \emph{stable in probability} if
    \[
        \lim_{x_0\to0}
        \mathbb{P}
        \left[
        \sup_{k\in\mathbb{N}}
        \|x_k\|>\epsilon
        \right]
        =0,
        \qquad \textrm{for every } \epsilon>0
    \]

    \item \emph{asymptotically stable in probability} if it is stable in probability and
    \[
        \lim_{x_0\to0}
        \mathbb{P}
        \left[
        \lim_{k\to\infty}
        \|x_k\|
        =0
        \right]
        =1;
    \]

    \item \emph{exponentially stable in probability} if there exists a constant $\gamma>1$, independent of $\omega$, such that
    \[
        \lim_{x_0\to0}
        \mathbb{P}
        \left[
        \lim_{k\to\infty}
        \|\gamma^k x_k\|
        =0
        \right]
        =1.
    \]
\end{enumerate}
For a set $\mathcal{Q}\subseteq\mathbb{R}^n$, the origin of \eqref{eq:stochastic_system} is said to be
\begin{enumerate}[label=(\roman*)]
    \item \emph{locally (globally) almost surely stable} in $\mathcal{Q}$ if, for every initial condition $x_0\in\mathcal{Q}$ (respectively, $x_0\in\mathbb{R}^n$), the sample paths remain in $\mathcal{Q}$ (respectively, $\mathbb{R}^n$) for all $k\ge0$ and converge to the origin almost surely;

    \item \emph{locally (globally) exponentially stable} in $\mathcal{Q}$ if it is locally (globally) almost surely stable and the convergence is exponential.
\end{enumerate}
\end{definition}

The following lemma, adapted from \cite{R20,R21}, recalls a sufficient condition for asymptotic convergence and Lyapunov stability in probability.

\begin{lemma}
\label{lem:stochastic_stability}
Consider the stochastic system \eqref{eq:stochastic_system}. Let $\{x_k\}_{k\ge0}$ be a Markov chain, and let $V:\mathbb{R}^n\rightarrow\mathbb{R}$ be a positive definite Lyapunov function. For some $\lambda>0$, define the set
\[
    \mathcal{Q}_\lambda
    \triangleq
    \{x_k:0\le V(x_k)<\lambda\}.
\]
Suppose that
\begin{align}\label{eq:lyapunov_difference}
    \mathbb{E}[V(x_{k+1})]-V(x_k)
    =
    -\vartheta(x_k)
    \le0,
    \qquad
    \forall k,
\end{align}
where $x_k\in\mathcal{Q}_\lambda$ and $\vartheta(\cdot)$ is continuous. Then the following statements hold.
\begin{enumerate}[label=(\roman*)]
    \item If $x_0\in\mathcal{Q}_\lambda$, then the sample paths remain in $\mathcal{Q}_\lambda$ with probability at least $1-V(x_0)/\lambda,$ the sequence $\{V(x_k)\}$ converges to a finite limit, and
    $\lim_{k\to\infty}\vartheta(x_k)=0$ almost surely.

    \item Suppose that the conditions in {\rm(i)} hold. If, for every $\gamma>0$, there exists $\delta>0$ such that
    $\vartheta(x_k)\ge\delta,$ whenever $\|x_k\|>\gamma$
    and $\vartheta(0)=0$, then the origin of \eqref{eq:stochastic_system} is globally almost surely stable.
\end{enumerate}
\end{lemma}

The adaptive quantized control law relies on acknowledgment messages that inform the controller whether the transmitted control packet is successfully received, i.e., the realization of $\eta(k)$. Following \cite{R15}, the time-varying nonlinear term $q_s(k,u)$ is incorporated into the controller update. For completeness, we summarize the controller design procedure below before stating the corresponding stability result.
\begin{enumerate}
    \item Select a positive definite matrix $R\in\mathbb{R}^{n\times n}$ and a scalar $\gamma\in(0,1)$.

    \item Solve the Riccati equation for $P\ge I_n$:
    \begin{align*}
        P
        =
        \Tilde{A}^\top P\Tilde{A}
        +
        R
        -
        \Tilde{A}^\top PB
        \left(B^\top PB\right)^{-1}
        B^\top P\Tilde{A}.
    \end{align*}

    \item Define
    \[
        A_s
        \triangleq
        \Tilde{A}
        +
        BK_g^{\prime\prime},
    \]
    where
    \[
        \Tilde{A}
        \triangleq
        A+BK_g^{\prime},
        \qquad
        K_g^{\prime\prime}
        \triangleq
        -
        (B^\top PB)^{-1}B^\top P\Tilde{A}.
    \]
    If $(A,B)$ is stabilizable, then $A_s$ is Hurwitz. Here, $\Tilde{A}$ is assumed to be known and unstable.

    \item Select $Q\in\mathbb{R}^{m\times m}$ and $\varepsilon>0$ such that
    $0<Q<2I_m,$ and
    \begin{align*}
        \frac{1}{\varepsilon}
        (2I_m-Q)
        -
        2B^\top PB
        \ge0.
    \end{align*}
    Such matrices $P$, $Q$, and scalar $\varepsilon$ always exist.
\end{enumerate}

The following theorem, adapted from \cite{R15}, summarizes the adaptive controller design and its stability guarantee.

\begin{theorem}
\label{thm:aqc}
Consider the uncertain discrete-time system \eqref{eq:plant}. Suppose that $A$ is unknown and satisfies $\sigma(A)<\bar{\sigma}_A$, the input matrix satisfies $\rho(B)=m$, and the pair $(A,B)$ is stabilizable. Assume further that acknowledgment messages are available, so that the controller knows whether packet loss has occurred. If the packet-loss bound $\bar{p}$ satisfies
\begin{align}\label{eq:packet_bound}
    \frac{\bar{p}}{1-\bar{p}}
    \left(
    \lambda_{\max}(P)\bar{\sigma}_A^2I_n-P
    \right)
    <\gamma R,
\end{align}
then, following the controller synthesis procedure above, the adaptive control law
\begin{align}\label{eq:adaptive_control}
    u(k)
    =
    2\left[M_1(k)+M_2(k)\right]^{-1}
    K(k)x(k),
\end{align}
where $K(k)\in\mathbb{R}^{m\times n}$ and the quantizer satisfies
\begin{align}\label{eq:sector_condition}
    R
    -
    2K^\top(k)
    \Delta(k)
    B^\top PB
    \Delta(k)
    K(k)
    \ge
    \gamma R
    >
    0,
\end{align}
for every $k\in\mathbb{N}_0$, together with the quantizer \eqref{eq:quantizer} and the gain update law
\begin{align}\label{eq:gain_update}
    K(k+1)
    =
    K(k)
    &-
    \frac{\eta(k)}
    {1+x^\top(k)Px(k)}
    QB^\dagger
    \left[
    x(k+1)
    \right.
    \nonumber\\
    &
    \left.
    -A_sx(k)
    -Bq_s(k,u(k))
    \right]
    x^\top(k),
\end{align}
guarantees Lyapunov stability. In particular, the equilibrium
\[
    (x(k),K(k))
    \equiv
    (0,K_g),
\]
where
\[
    K_g
    \triangleq
    -(B^\top PB)^{-1}B^\top PA,
    \qquad
    K_g
    =
    K_g'+K_g'',
\]
is asymptotically stable, and $\lim_{k\to\infty}x(k)=0,$ for all $x_0\in\mathbb{R}^n.$
\end{theorem}

Condition \eqref{eq:sector_condition} must hold for every time step. Consequently, the sector bounds $M_1(k)$ and $M_2(k)$ are also required to vary with time. A simple choice is to fix
$M_1(k)\equiv I_m,$ and define
\[
    M_2(k)\in\{1+\hat{a}\mu_i^j:j\in\mathbb{I}\},
\]
where $\hat{a}>0$ and $\mu_i>0$ for every $i$. Since $M_2(k)-M_1(k)\succ0$, it follows that $M_2(k)>I_m$. Assume further that $M_2(k)$ is uniformly bounded by a constant $M_\phi$, that is,
\[
    M_1(k)<M_2(k)\le M_\phi.
\]
The remaining task is to determine the appropriate index $j$ at each time step so that \eqref{eq:sector_condition} is satisfied. The resulting adaptive quantized control algorithm is summarized in Algorithm~\ref{alg1}. In the next section, we introduce the reinforcement learning controller for comparison.

\begin{algorithm}[t!]
    \caption{Adaptive Quantized Control}\label{alg:aqc}
    \begin{algorithmic}[1]
    \Require $A,B,x_0,k_{\rm on},K_0,R\coloneqq I_2\rightarrow P\ge I_2,\bar p$
    \State Choose $Q$ such that $0<Q<2I_m$
    \State $K_g \leftarrow -(B^\top PB)^{-1}B^\top PA$
    \State $A_s \leftarrow A+BK_g$
    \For{$k=1:t_{\max}$}
        \State Generate $p_c$ and set $\eta(k)\leftarrow 1$ if $p_c>\bar p$; otherwise, set $\eta(k)\leftarrow0$
        \If{$k\ge t_d$} \Comment{$t_d$ denotes the switching time}
            \State $(A,B)\leftarrow(A_w,B_w)$
        \EndIf
        \State Choose $\gamma\in(0,1)$ and compute $\Delta(k)$ from \eqref{eq:delta}
        \If{$\delta\ge1$} \Comment{enforce $0<\delta<1$}
            \State $M_2(k)\leftarrow M_\phi$, $\rho\leftarrow 1/\phi$
        \Else
            \For{$j_i$, $i=1,\ldots,n$} \Comment{$M_1<M_2(j_i)<M_\phi$}
                \If{$\delta>\frac{1}{\beta}(M_2+M_1)^{-1}(M_2-M_1)$}
                    \State $j\leftarrow j_i$ and \textbf{stop}
                \EndIf
            \EndFor
            \State $u(k)\leftarrow$ \eqref{eq:adaptive_control}, \quad $v(k)\leftarrow$ \eqref{eq:log_quantizer}
        \EndIf
        \State $x(k+1)\leftarrow Ax(k)+\eta(k)Bv(k)$
        \State $q_s(k,u(k))\leftarrow$ \eqref{eq:quantizer_decomp}
        \State $K(k+1)\leftarrow$ \eqref{eq:gain_update}
    \EndFor
    \end{algorithmic}
\end{algorithm}

\section{Deep Deterministic Policy Gradient Reinforcement Learning}
\label{sec:ddpg}

We briefly review the deep deterministic policy gradient (DDPG) algorithm \cite{R11,R18}, which serves as the reinforcement learning benchmark throughout this paper. Unlike stochastic policy methods, DDPG employs a deterministic policy
\[
    u=\mu_\theta(x),
\]
which maps the state directly to the control action, instead of the stochastic policy
$\pi_\theta(u|x)=\mathbb{P}[u|x;\theta].$
The state and action spaces are denoted by $x\in\mathcal{S}$ and $u\in\mathcal{A}$, respectively.

The learning problem is formulated as a Markov decision process (MDP) defined by the tuple $(\mathcal{S},\mathcal{A},p,r)$, where $p_1(x_1)$ denotes the initial-state distribution and
$p(x_{t+1}|x_t,u_t)\coloneqq p(x_{t+1}|x_1,u_1,\ldots,x_t,u_t)$
is the stationary state-transition probability. DDPG is an off-policy actor--critic algorithm in which the critic approximates the action-value function, while the actor learns the policy by maximizing the expected return. The objective is to maximize
\begin{align}\label{eq:objective}
    J(\mu_\theta)
    =
    \int_{\mathcal{S}}
    \rho^\mu(x)
    \int_{\mathcal{A}}
    \mu_\theta(x,u)
    r(x,u)
    \,du\,dx,
\end{align}
where $\rho^\mu(x)$ denotes the discounted state distribution induced by the policy $\mu_\theta$. The corresponding discounted return is
\begin{align}\label{eq:return}
    r_t^\gamma
    =
    \sum\nolimits_{k=t}^{\infty}
    \gamma^{k-t}
    r(x_k,u_k),
\end{align}
where $\gamma\in(0,1)$ is the discount factor. An optimal policy $\mu_\theta^\ast$ satisfies
$J(\mu_\theta^\ast)\ge J(\mu_\theta)$ for every admissible policy $\mu_\theta$.

\begin{figure}[t!]
    \centering
    \includegraphics[width=.5\linewidth]{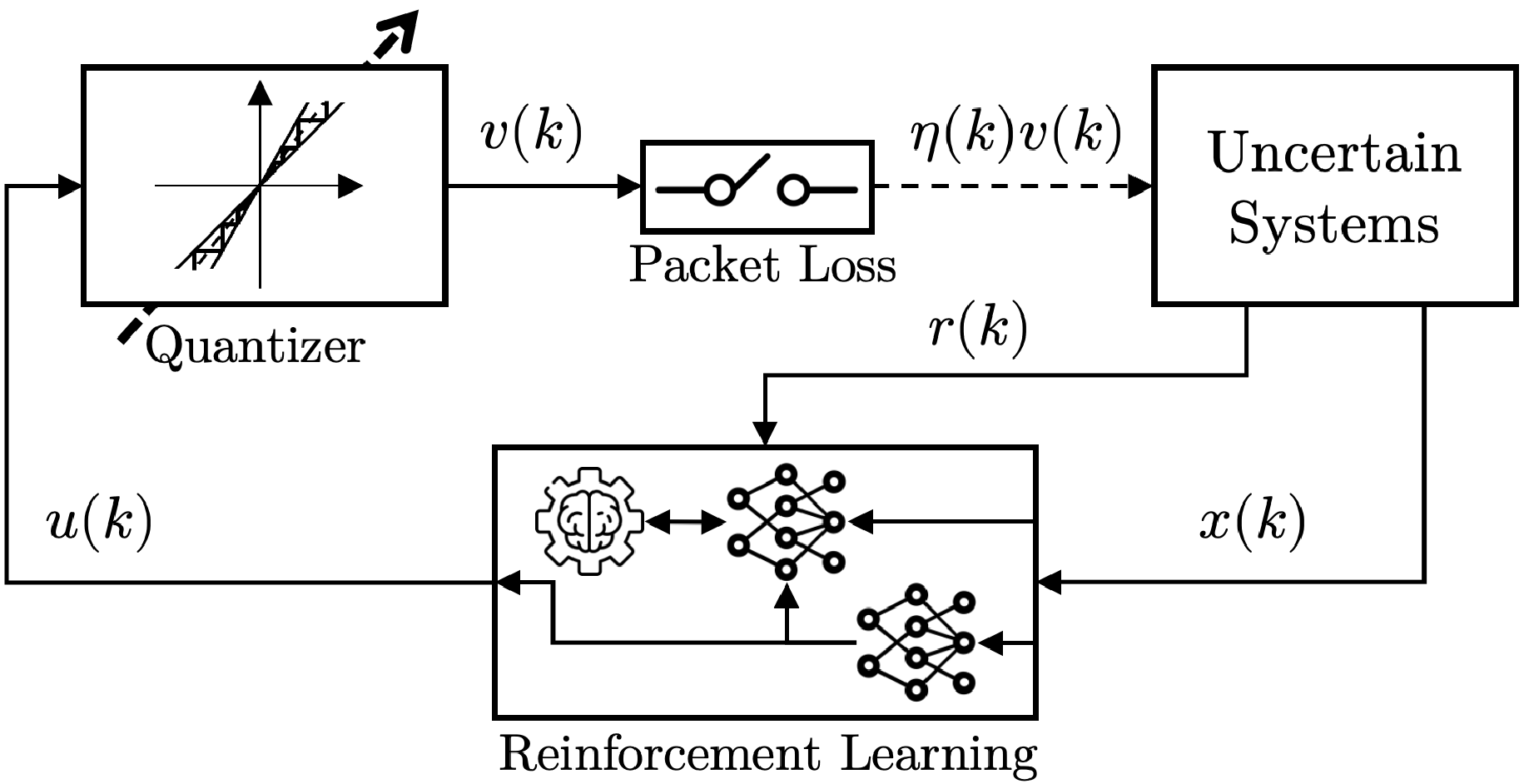}
    \caption{Actor--critic DDPG controller with the time-varying quantizer $q(k,u(k))$. The arrows have the same interpretation as those in Fig.~\ref{F1}.}
    \label{fig:ddpg_structure}
\end{figure}

DDPG is closely related to the deep Q-network (DQN), with the main difference being that DDPG is designed for continuous action spaces. The state--action value function $Q^\mu(x,u)$ represents the expected return under the policy $\mu_\theta$. The corresponding optimal value function is denoted by $Q^\ast(x,u)$, from which the optimal policy is obtained as
\[
    u^\ast(x)=\arg\max_u Q^\ast(x,u).
\]
Unlike DQN, which searches over a finite action set, DDPG assumes that the action space is continuous and that $Q^\ast(x,u)$ is differentiable with respect to the action variable. Consequently, the maximization
\[
    \max_u Q(x,u)
\]
is approximated by $Q(x,\mu_\theta(x)),$
allowing the policy to be updated using gradient-based optimization. The corresponding Bellman equation is
\begin{align}\label{eq:bellman}
    Q^\ast(x,u)
    =
    \mathbb{E}_{x^\prime\sim\mathbb{P}}
    \left[
        r(x,u)
        +
        \gamma
        \max_{u^\prime}
        Q^\ast(x^\prime,u^\prime)
    \right],
\end{align}
where $x^\prime\sim\mathbb{P}$ denotes the next state sampled from the transition probability $\mathbb{P}(\cdot|x,u)$.

In practice, the action--value function is approximated by a neural network $Q_\phi(x,u)$ with parameters $\phi$. The network is trained using a replay buffer
$\Phi=\{(x,u,r,x^\prime,d)\},$
where $d$ indicates whether $x^\prime$ is a terminal state. The target value and the corresponding mean squared Bellman error (MSBE) are given by
\begin{align}\label{eq:msbe}
\begin{aligned}
    \xi_\Phi
    &=
    r
    +
    \gamma(1-d)
    \max_{u^\prime}
    Q^\ast(x^\prime,u^\prime),\\
    L(\phi,\Phi)
    &=
    \mathbb{E}_{x,u,r,x^\prime,d\sim\Phi}
    \!\left[
        \left(
            Q_\phi(x,u)-\xi_\Phi
        \right)^2
    \right].
\end{aligned}
\end{align}
The reward is designed as a function of the tracking error $e\triangleq x$. Introducing a small positive constant $\psi$ to avoid division by zero and a threshold $T_h$, the reward function is defined as
\begin{align}\label{eq:reward}
    r(k)=
    \begin{cases}
        -100,
        & |x|>T_h,\\[1mm]
        \dfrac{1}{e+\psi},
        & \text{otherwise}.
    \end{cases}
\end{align}

Recall the uncertain linear system \eqref{eq:plant} together with the DDPG controller shown in Fig.~\ref{fig:ddpg_structure}. The state $x(k)\in\mathbb{R}^{n_p}$
is also taken as the measured output, i.e., $y(k)\coloneqq x(k),$
while $u(k)\in\mathbb{R}^{n_u}$
denotes the control input.

\begin{figure*}[t!]
    \centering
    \includegraphics[width=.95\linewidth]{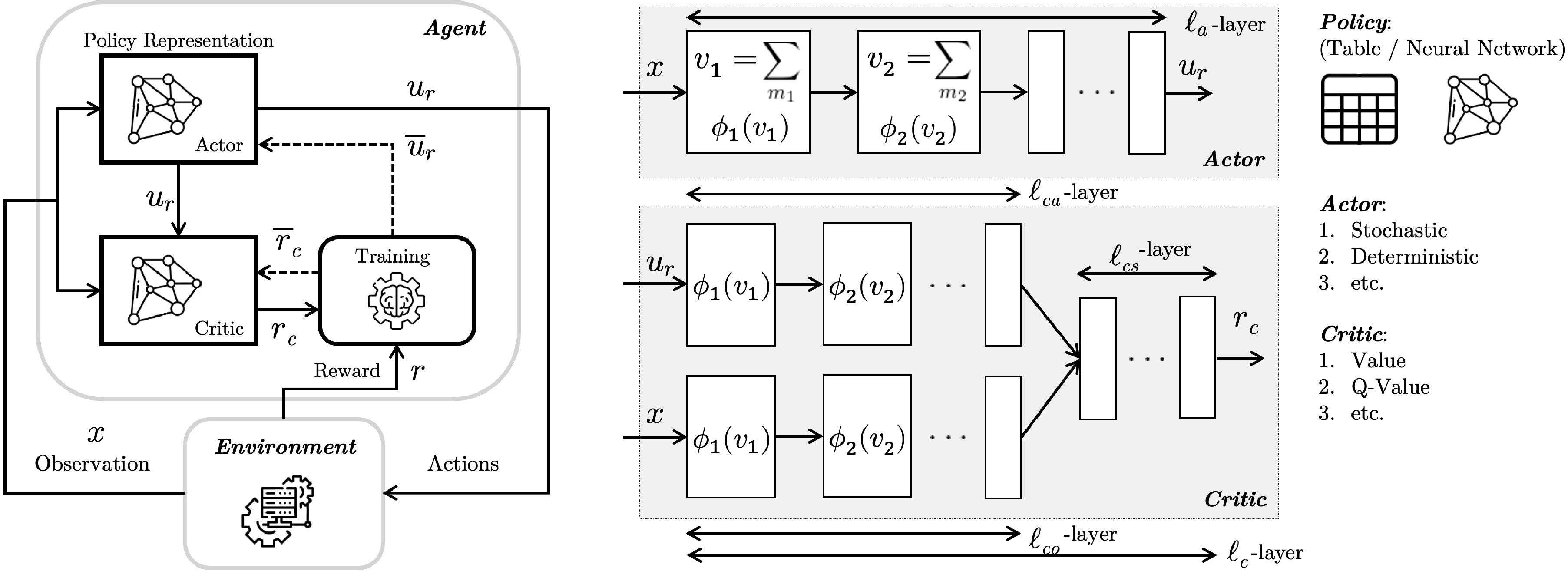}
    \caption{Architecture of the actor--critic DDPG controller.}
    \label{fig:ddpg_design}
\end{figure*}

The DDPG controller consists of two feed-forward neural networks (FFNNs): an \emph{actor} network that maps the measured state to the control input, and a \emph{critic} network that estimates the corresponding action-value function. As illustrated in Fig.~\ref{fig:ddpg_design}, the actor consists of $\ell_a$ hidden layers, whereas the critic receives both the state and action as inputs. The state and action paths comprise $\ell_{co}$ and $\ell_{ca}$ layers, respectively, followed by $\ell_{cs}$ common layers. Consequently, the total number of critic layers is
\[
    \ell_c
    \triangleq
    \max(\ell_{co},\ell_{ca})
    +
    \ell_{cs}.
\]

The actor network is described by
\begin{subequations}\label{eq:actor}
\begin{align}
    \phi_0(k)
    &=x(k), \label{eq:actor_input}\\
    \phi_i(k)
    &=\Delta_i\!\left(
        W_i\phi_{i-1}(k)+b_i
    \right)
    \triangleq
    \Delta_i(v_i(k)),
    \qquad
    i=1,\ldots,\ell_a,
    \label{eq:actor_hidden}\\
    u^{(n_u)}(k)
    &=W_{\ell_a+1}\phi_{\ell_a}(k)
    +b_{\ell_a+1}
    \triangleq
    v_{\ell_a+1}^{(n_u)}.
    \label{eq:actor_output}
\end{align}
\end{subequations}

For the $i$-th hidden layer,
$W_i\in\mathbb{R}^{m_i\times m_{i-1}}$
and
$b_i\in\mathbb{R}^{m_i}$
denote the weight matrix and bias vector, respectively, where $m_0=n_p$. The corresponding pre-activation vector is given by
\begin{align}\label{eq:preactivation}
    v_i^{(j)}(k)
    =
    \sum_{t=1}^{m_{i-1}}
    W_i^{(j,t)}
    \phi_{i-1}^{(t)}(k)
    +
    b_i^{(j)},
    \qquad
    j=1,\ldots,m_i.
\end{align}

The activation function
$\Delta_i$
is applied elementwise according to
\[
    \Delta_i(v_i)
    =
    \bigl[
    \lambda(v_1),
    \ldots,
    \lambda(v_{m_i})
    \bigr]^\top,
\]
where $\lambda(\cdot)$ denotes either the hyperbolic tangent or the ReLU activation function. The output layer has dimension $n_u$, matching the dimension of the control input.

For subsequent analysis, define
\[
    v_q
    =
    \begin{bmatrix}
    v_1^\top &
    \cdots &
    v_{\ell_a}^\top
    \end{bmatrix}^\top,
    \qquad
    \phi_q
    =
    \begin{bmatrix}
    \phi_1^\top &
    \cdots &
    \phi_{\ell_a}^\top
    \end{bmatrix}^\top,
\]
where
\[
    n_q
    =
    \sum\nolimits_{i=1}^{\ell_a}
    n_i.
\]
The vectors $v_q$ and $\phi_q$ collect the pre-activation and post-activation variables of all hidden layers, respectively, and the resulting control signal $u(k)$ is applied to the plant and subsequently supplied to the critic network.

The actor network admits the compact input--output representation
\begin{align}\label{eq:actor_compact}
    \begin{bmatrix}
         u(k)\\
         v_q(k)
    \end{bmatrix}
    =
    N
    \begin{bmatrix}
         x(k)\\
         \phi_q(k)\\
         1
    \end{bmatrix},
    \qquad
    \phi_q(k)=\Delta(v_q(k)),
\end{align}
where $N$ is constructed from the weight matrices and bias vectors of the actor network, namely
$W_q\in\mathbb{R}^{n_q+1}$ and
$b_q\in\mathbb{R}^{n_q+1}$.
Specifically,
{\small
\begin{align}\label{eq:actor_matrix}
    N
    \triangleq
    \left[
    \begin{array}{c|cccc|c}
        0 & 0 & 0 & \cdots & W_{\ell_a+1} & b_{\ell_a+1}\\
        \hline
        W_1 & 0 & \cdots & 0 & 0 & b_1\\
        0 & W_2 & \cdots & 0 & 0 & b_2\\
        \vdots & \vdots & \ddots & \vdots & \vdots & \vdots\\
        0 & 0 & \cdots & W_{\ell_a} & 0 & b_{\ell_a}
    \end{array}
    \right].
\end{align}
}

The representation \eqref{eq:actor_compact} separates the linear and nonlinear components of the FFNN, whose stability properties are studied in \cite{R22}. The critic network is constructed in the same manner as the actor network and is given by
\begin{align}\label{eq:critic}
\begin{aligned}
    \varphi_0(t)
    &=
    \begin{bmatrix}
        u(t)\\
        x(t)
    \end{bmatrix},
    \\[1mm]
    \varphi_i^{u}(t)
    &=
    \Delta_i^{u}
    \left(
        W_i^{u}\varphi_{i-1}^{u}(t)+b_i^{u}
    \right),
    \qquad
    i=1,\ldots,\ell_{ca},
    \\[1mm]
    \varphi_i^{x}(t)
    &=
    \Delta_i^{x}
    \left(
        W_i^{x}\varphi_{i-1}^{x}(t)+b_i^{x}
    \right),
    \qquad
    i=1,\ldots,\ell_{co},
    \\[1mm]
    \varphi_{\ell_0}(t)
    &=
    \varphi_{\ell_{ca}}^{u}(t)
    +
    \varphi_{\ell_{co}}^{x}(t),
    \qquad
    \ell_0
    \coloneqq
    \max(\ell_{ca},\ell_{co}),
    \\[1mm]
    \varphi_j(t)
    &=
    \Delta_j
    \left(
        W_j\varphi_{j-1}(t)+b_j
    \right),
    \qquad
    j=\ell_0+1,\ldots,\ell_0+\ell_{cs},
    \\[1mm]
    r_c(t)
    &=
    W_{\ell_0+\ell_{cs}+1}
    \varphi_{\ell_0+\ell_{cs}}(t)
    +
    b_{\ell_0+\ell_{cs}+1}
    \coloneqq
    v_{\ell_0+\ell_{cs}+1}
    \in\mathbb{R}^{r_c}.
\end{aligned}
\end{align}
where the notation follows directly from the actor network in
\eqref{eq:actor}--\eqref{eq:actor_matrix}. To distinguish the reinforcement learning control input from the adaptive quantized controller, we denote the actor output by
\[
    u_r(k)\coloneqq u(k).
\]
Suppose that $(x^\ast,u_r^\ast)$ is an equilibrium of the closed-loop system. Propagating the equilibrium state through the actor and critic networks yields the corresponding equilibrium values $(v^\ast,\phi^\ast)$ of the activation inputs and outputs. Consequently,
$(x^\ast,u_r^\ast,v^\ast,\phi^\ast)$
is an equilibrium of the interconnected system consisting of
\eqref{eq:plant},
\eqref{eq:actor}, and
\eqref{eq:critic},
provided that $x^\ast=Ax^\ast+Bq(k,u_r^\ast).$

Algorithm~\ref{alg:rl_quantizer} summarizes the proposed reinforcement learning controller with the time-varying quantizer. During training, the controller is learned using the nominal system matrix $A$. The learned policy is subsequently evaluated under plant variations by replacing $A$ with $A_w$ and introducing packet loss. This setting provides a direct comparison with the adaptive quantized controller, which is designed for uncertain systems through the stabilizing gain $K_g$ satisfying
\[
    A_s=A+BK_g.
\]

\begin{algorithm}[t!]
\caption{Reinforcement Learning with Quantizer}\label{alg:rl_quantizer}
\begin{algorithmic}[1]
\Require $A,B,x_0,k_{\rm on},R\coloneqq I_2\rightarrow P\ge I_2$, FFNN parameters $\psi$, learning rates $\tau_a,\tau_c$, discount factor $\gamma$, sampling time $t_s$, number of episodes $\mathcal{E}$, and terminal time $\mathcal{T}$
\For{$k=1:t_{\max}$}
    \State Observe the state $x(k)$
    \State Compute the actor output $u_r(k)$
    \State Observe the reward $r(k)$
    \If{$k\ge t_d$} \Comment{$t_d$ denotes the switching time}
        \State $(A,B)\leftarrow(A_w,B_w)$
    \EndIf
    \State Choose $\gamma\in(0,1)$
    \State Compute $\Delta(k)$ from \eqref{eq:delta} with $K(k)x(k)$ replaced by $u_r(k)$
    \State Compute the quantized input $v(k)=q(k,u_r(k))$ using \eqref{eq:log_quantizer}
    \State Update the state:
    \[
        x(k+1)=Ax(k)+\eta(k)Bv(k)
    \]
\EndFor
\end{algorithmic}
\end{algorithm}

\section{Numerical Results and Findings}\label{sec:results}

In this section, we compare the performance of the adaptive quantized control (AQC) and deep deterministic policy gradient (DDPG) controllers under four simulation scenarios:
\begin{enumerate}[label=(\roman*)]
    \item AQC without packet loss;
    \item AQC with packet loss satisfying $\bar{p}_1=0.15$, $\bar{p}_2=0.30$, and $\bar{p}_3>0.5$;
    \item DDPG with the input quantizer;
    \item DDPG with the input quantizer under the same packet-loss conditions as in (ii).
\end{enumerate}
For each scenario, the system is switched from the nominal dynamics $A$ to the perturbed dynamics $A_w$ at
\[
    t_d\in\{10,15,20,25,30\}\ {\rm s},
\]
while the controller is activated at $k_{\rm on}=1\ {\rm s}.$

Consider the uncertain second-order system subject to packet loss,
\begin{align}\label{eq:simulation_system}
    \begin{aligned}
        z(k+2)
        +\beta_\zeta z(k+1)
        +\alpha_\zeta z(k)
        &=
        \eta(k)b_\zeta v(k),\\
        z(0)&=z_0,\qquad
        z(1)=z_1,
    \end{aligned}
\end{align}
where $k\in\mathbb{N}_0$. The parameters $\alpha_\zeta,\beta_\zeta\in\mathbb{R}$, $\zeta\in\{1,2\}$, are unknown system coefficients, while $b_\zeta\in\mathbb{R}$ is a known input gain. The control input is denoted by $v(k)\in\mathbb{R}$ and is generated through the quantizer described in Section~\ref{sec:aqc}. Packet transmission is governed by the binary random variable $\eta(k)$ satisfying
$\mathbb{P}\{\eta(k)=0\}\le\bar{p}_i,$ where $\bar{p}_i$ denotes the packet-loss bound.

Defining the state variables $x_1(k)=z(k)$ and $x_2(k)=z(k+1)$.
The nominal and perturbed system matrices are
\[
    A=
    \begin{bmatrix}
    0 & 1\\
    -\alpha_1 & -\beta_1
    \end{bmatrix},
    \qquad
    A_w=
    \begin{bmatrix}
    0 & 1\\
    -\alpha_2 & -\beta_2
    \end{bmatrix},
\]
with $x=[x_1,x_2]^\top$, $x(0)=x_0$, and $B = [0,b_\zeta]^\top,\forall \zeta$.
For the AQC design, we choose $R=I_2,$ which yields the Riccati solution
$P=\diag(1,2)>I_2.$ The system parameters are selected as
\[
    \alpha_1=1.5,\qquad
    \alpha_2=2,\qquad
    \beta_1=\beta_2=-0.5,\qquad
    b_\zeta=0.4.
\]
Since only the first state is quantized, we fix $M_1(k)\equiv1,$ and choose
\[
    M_2(k)\in
    \left\{
    1+3(1.3)^j:
    j\in\mathbb{I}
    \right\}.
\]
The index $j$ determines the time-varying quantization level, satisfying
\[
1=M_1<M_2(k)<M_\phi,
\qquad
M_\phi=10.
\]

\begin{figure}[t!]
    \centering
    \subfloat[\label{F5c} Adaptive Quantized Control, $\eta(k)=1,\forall k$]{\includegraphics[width=.455\textwidth]{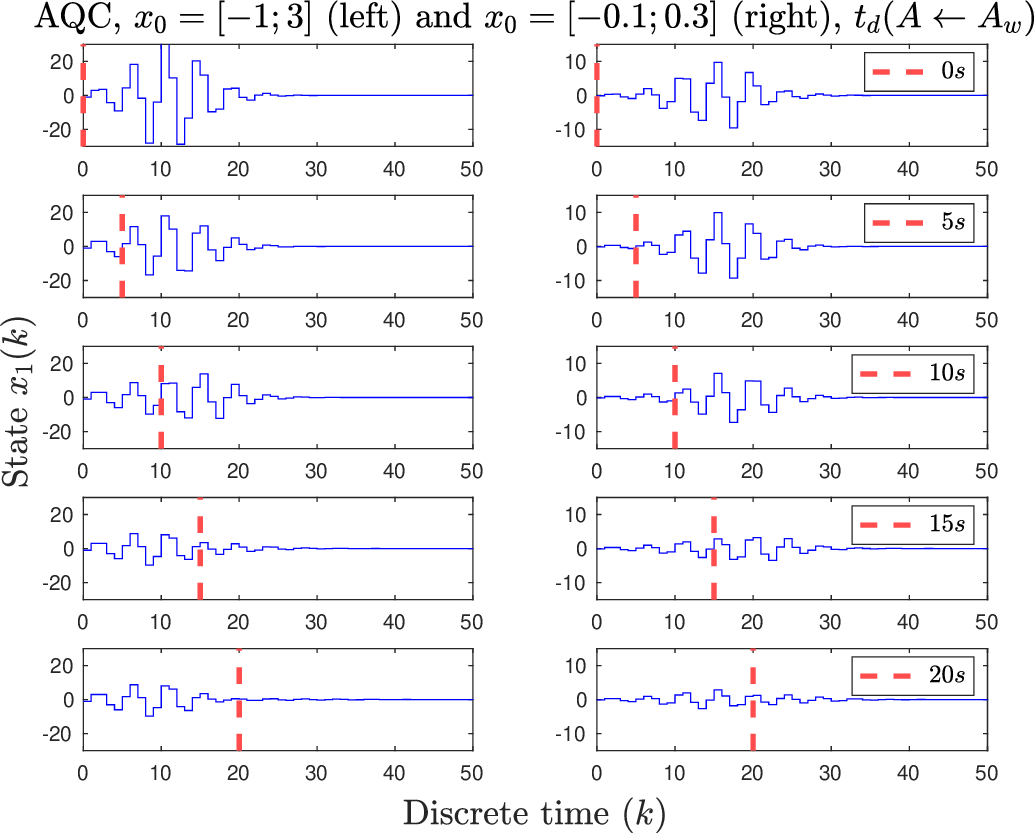}} \qquad
    \subfloat[\label{F5f} Reinforcement Learning, $\eta(k)=1,\forall k$]{\includegraphics[width=.45\textwidth]{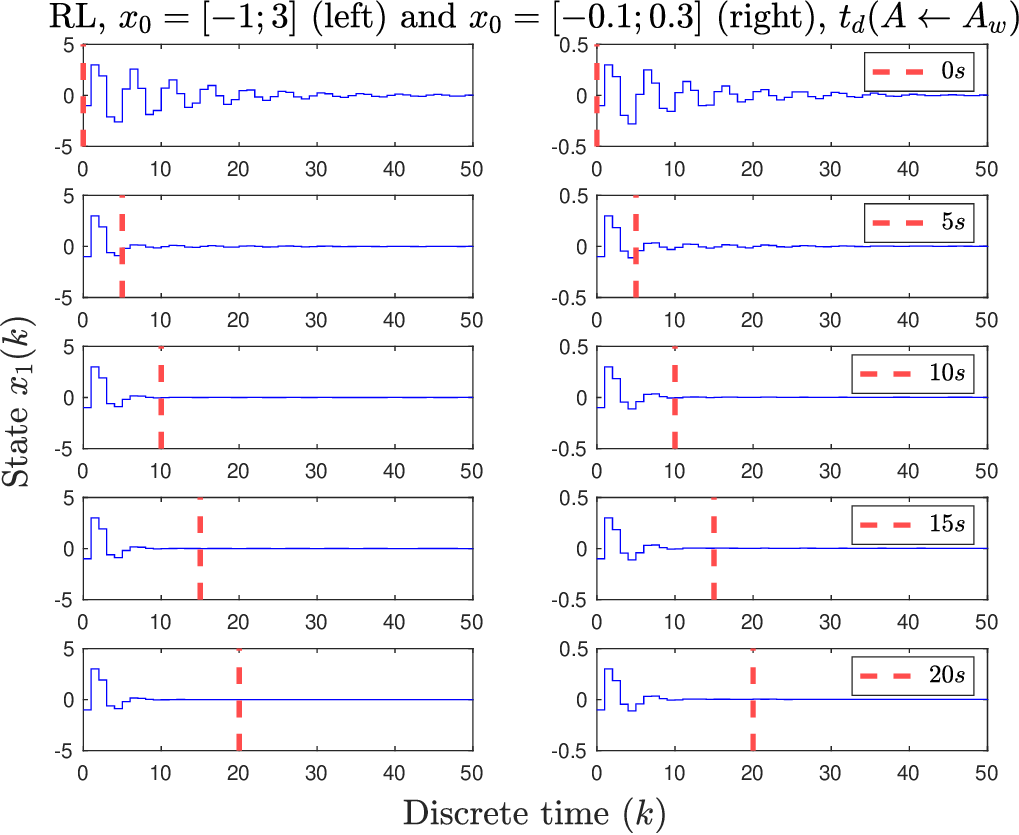}}
    \caption{\small Performances of AQC and DRL with two different initial condition $x_0$ and various dynamic change time $t_d$.}%
    \label{F5}%
\end{figure}
\begin{figure*}[t!]
    \centering
    \subfloat[\label{F6a}]{\includegraphics[width=.324\linewidth]{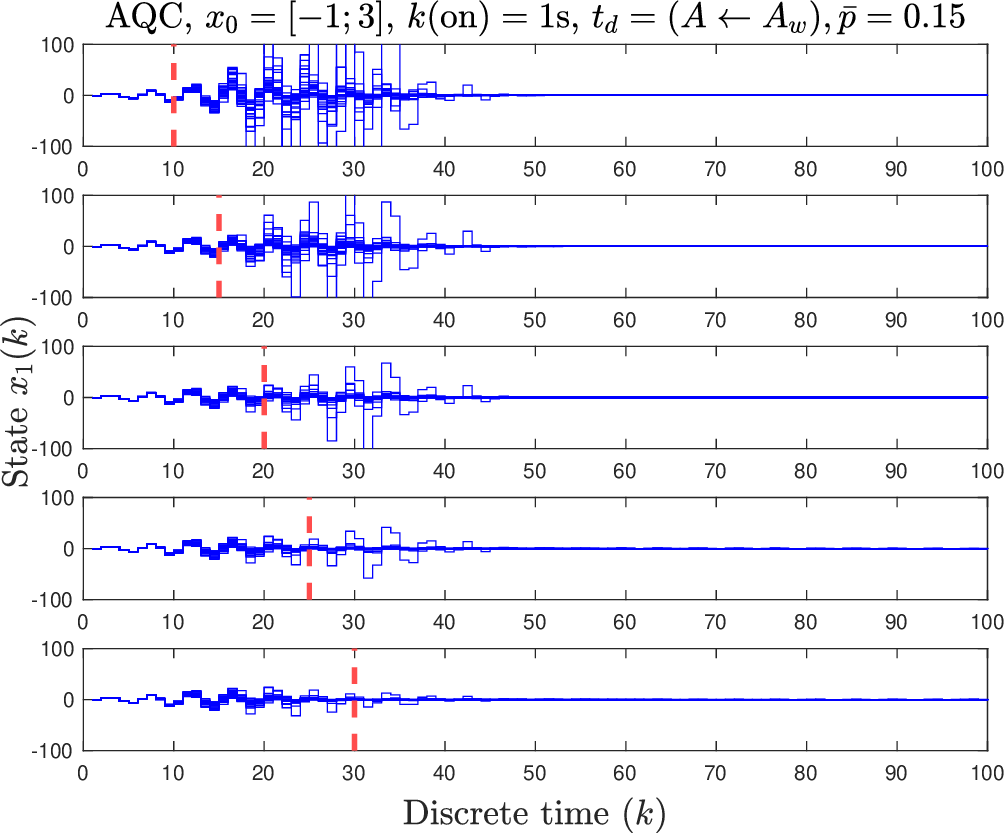}}\;
    \subfloat[\label{F6b}]{\includegraphics[width=.324\linewidth]{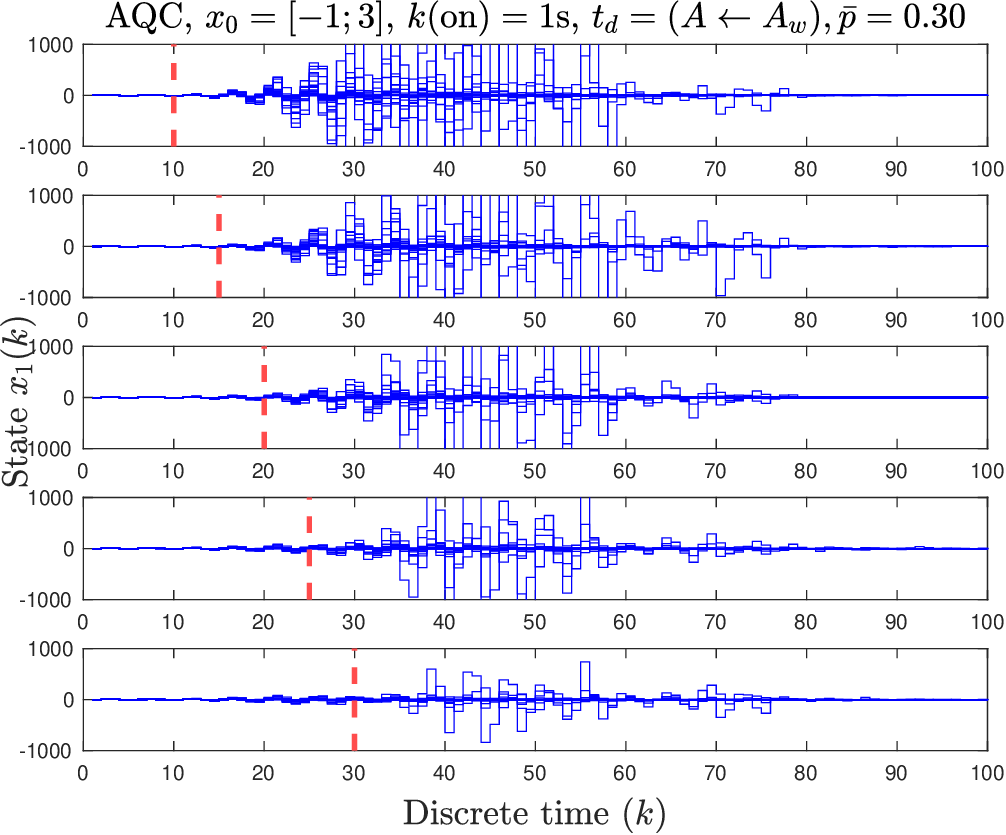}}\;
    \subfloat[\label{F6c}]{\includegraphics[width=.328\linewidth]{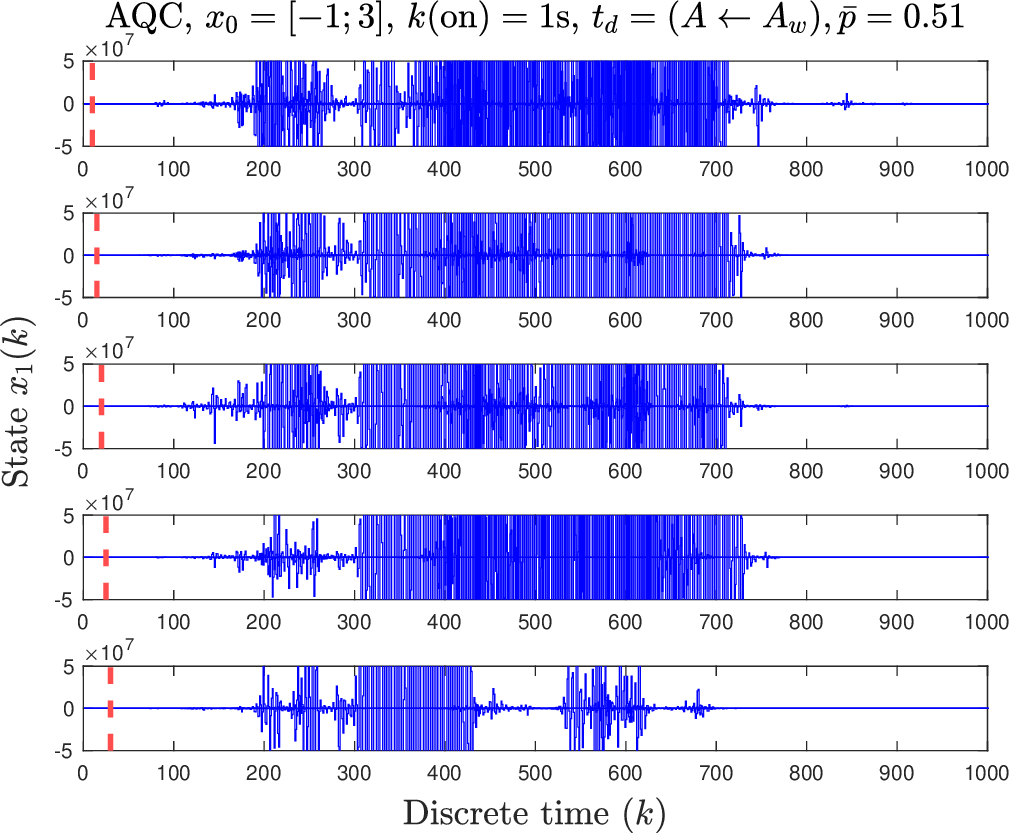}}\\
    \subfloat[\label{F6d}]{\includegraphics[width=.324\linewidth]{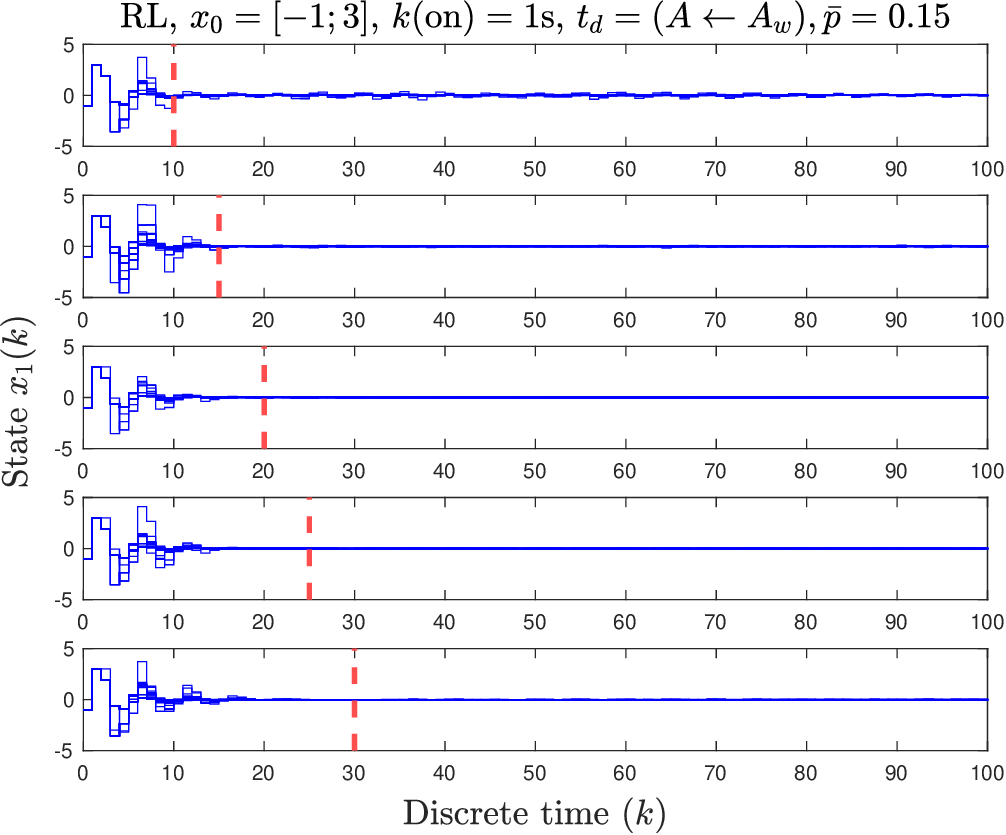}}\;
    \subfloat[\label{F6e}]{\includegraphics[width=.324\linewidth]{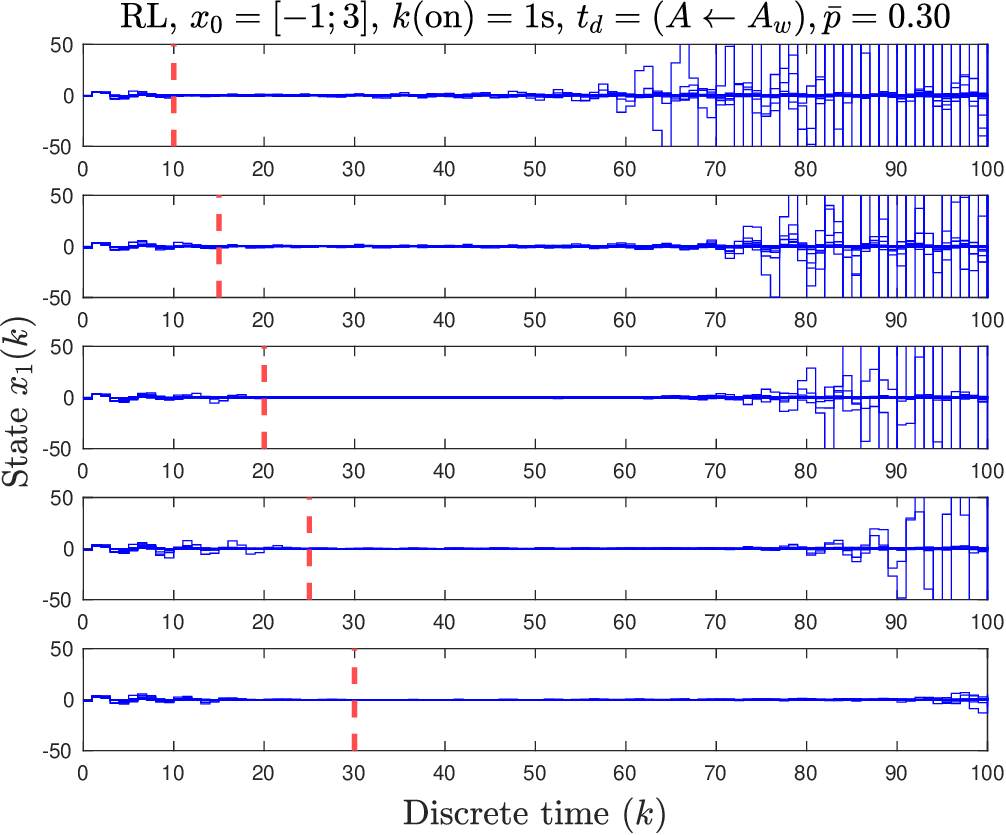}}\;
    \subfloat[\label{F6f}]{\includegraphics[width=.328\linewidth]{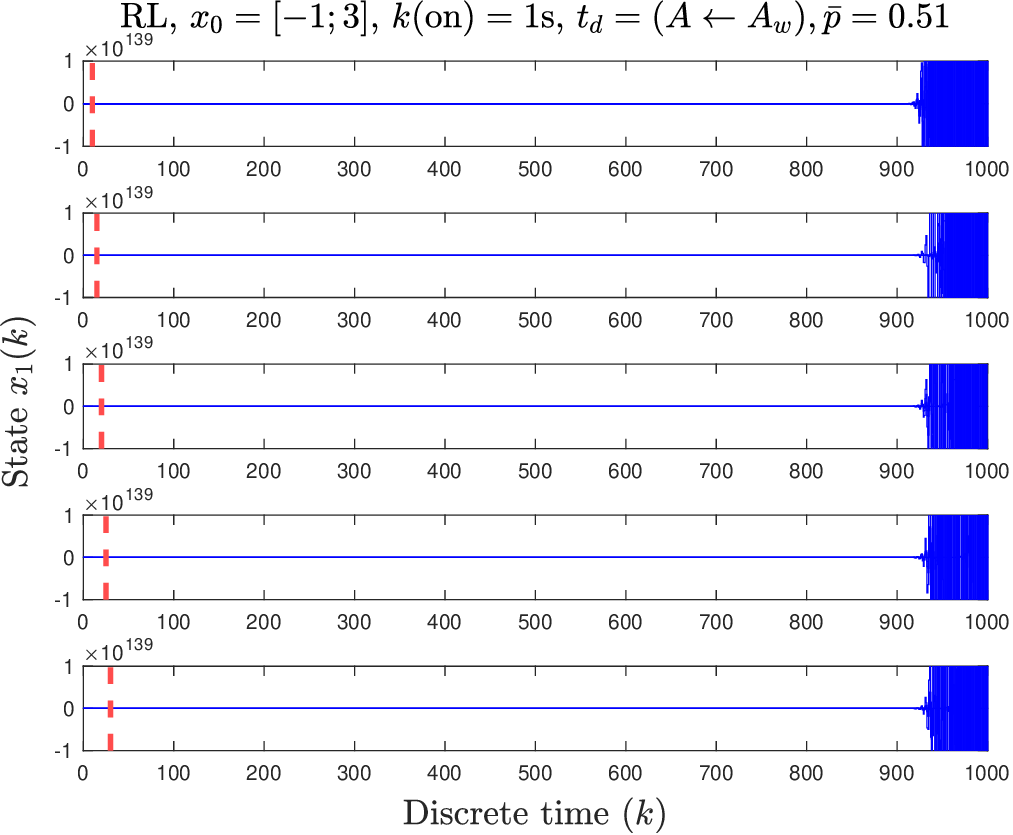}}
    \caption{\small Comparison of AQC and DRL under varying levels of packet loss $\eta(k)$.}%
    \label{F6}%
\end{figure*}

The DDPG controller employs the same time-varying quantizer as the AQC, with the control signal generated by the trained actor network. The observation is given by
\[
    e(k)=x(k)=y(k),
\]
while the action space is bounded by $|\mathcal{A}|<\mathcal{A}_\phi\coloneqq100.$

The learning rates of the actor and critic networks are chosen as $\tau_a=10^{-4}$ and $\tau_c=10^{-3}$, respectively. The actor network consists of 25 neurons with \texttt{tanh} and \texttt{softmax} activation functions, whereas the critic network employs ReLU activation functions with 50 and 25 neurons for the state and action branches, respectively. Finally, the remaining training parameters are selected as
\[
    \gamma=0.9,\qquad
    t_s=1~{\rm s},\qquad
    \mathcal{E}=1000,\qquad
    \mathcal{T}=750.
\]

The simulation results are presented in Figs.~\ref{F5} and \ref{F6}. Each simulation considers a dynamic switch from the nominal system $A$ to the perturbed system $A_w$ at the prescribed switching time $t_d$ for three different initial conditions. Figure~\ref{F5} shows that the AQC successfully stabilizes the system under all switching scenarios. As expected, earlier switching times produce larger transient responses since the adaptive gain $K(k)$ has had less time to converge toward its steady-state value.

Figure~\ref{F6} illustrates the effect of packet loss. Since the AQC incorporates acknowledgment messages, the controller is informed whenever a packet is lost through the nonlinear term $q_s(k,u(k))$. Consequently, the control input $v(k)$ and the corresponding quantizer parameters are continuously updated, even when $\eta(k)=0$. As shown in \cite{R15}, Lyapunov stability is guaranteed whenever the packet-loss bound satisfies \eqref{eq:packet_bound}. Without acknowledgment messages, this guarantee is restricted to the range
\[
    0\le\bar{p}\le0.5.
\]

The DDPG controller is trained using the nominal system $A$ together with the same quantizer shown in Fig.~\ref{F2}. As observed in Fig.~\ref{F5}, the trained policy remains effective under moderate dynamic changes and successfully stabilizes the perturbed system for sufficiently small switching times. The initial transient is mainly due to the mismatch between the training initial condition $x_0=[0,0]^\top$ and the initial conditions used during testing.

When packet loss is introduced, however, the AQC exhibits superior robustness compared with the DDPG controller. This difference is expected since the AQC is equipped with explicit stability guarantees for uncertain systems with packet loss, whereas the DDPG controller is trained only within the nominal environment and must generalize beyond its training distribution. It is also worth emphasizing that the AQC is designed using the stabilizing matrix $A_s=A+BK_g$, while the DDPG controller is trained using the nominal system matrix $A$ before being evaluated on the perturbed system $A_w$.

\begin{remark}
    Table~\ref{T1} summarizes the stability performance of the three controllers for different values of $\alpha_2$, where the system switches from $A$ to $A_w$ at time $t_d$. The DDPG controller performs well only within a neighborhood of the training environment ($\alpha_2=1$), whereas the AQC remains stable over a significantly wider range of parameter variations. As expected, the adaptive controller (AC) exhibits the strongest robustness and remains stable for all values of $\alpha_2$ considered in Table~\ref{T1}.
\end{remark}

\begin{table}[t!]
    \centering
    \small
    \begin{tabular}{|c|ccccccccc|} \hline
        $\alpha_2 = $ & 1 & 2 & 3 & 4 & 5 & $\dots$ & 18 & 19 & 20 \\\hline
        RL & \greencheck & \greencheck & $\redtimes$ & $\redtimes$ & $\redtimes$ & $\dots$ & $\redtimes$ & $\redtimes$ & $\redtimes$\\ 
        AQC & \greencheck & \greencheck & \greencheck & \greencheck & \greencheck & $\dots$ & \greencheck & $\redtimes$ & $\redtimes$\\
        AC & \greencheck & \greencheck & \greencheck & \greencheck & \greencheck & $\dots$ & \greencheck & \greencheck & \greencheck\\ \hline 
    \end{tabular}
    \caption{Stability comparison of three control methods across different dynamic switches $A\rightarrow A_w$ at time $t_d$}
    \label{T1}
\end{table}

\section{Conclusion}\label{sec:conclusion}

This paper presented a comparative study between adaptive quantized control (AQC) and deep deterministic policy gradient (DDPG) reinforcement learning for uncertain linear systems with input quantization, packet loss, and dynamic model switching. Numerical results showed that the DDPG controller achieves improved transient performance within the training environment, whereas the AQC exhibits superior robustness in the presence of model uncertainty and packet loss due to its explicit Lyapunov stability guarantees. Overall, the results highlight the complementary strengths of adaptive control and reinforcement learning, while illustrating the limitations of learning-based controllers when operating beyond the training environment.

Future research will focus on extending the proposed framework to networked multi-agent systems, where communication delays, packet losses, disturbances, and distributed interactions play a fundamental role. In particular, we are interested in developing reinforcement learning algorithms with rigorous stability guarantees for cooperative control, fault-tolerant coordination, and distributed adaptive control over communication networks, building upon our recent studies in adaptive and resilient multi-agent control \cite{Wafi-MRAC,Wafi-LCSS24,Yang2022,Wafi-ASCC26-FTC,Liu2024,Wafi-ASCC26-DNC,Gao2022}.

\bibliographystyle{ieeetr}
\bibliography{bibliography}

\end{document}